\documentclass[12pt]{amsart}
\usepackage{amsmath,amssymb,latexsym,amsthm,graphicx,cite}
\newtheorem{theorem}{Theorem}
\newtheorem{lemma}[theorem]{Lemma}

\newtheorem{corollary}[theorem]{Corollary}
\newtheorem{definition}[theorem]{Definition}
\def\qed{$\Box$}

\def\R{\Bbb{R}}

\def\RR{\Bbb{R}}
\def\ZZ{\Bbb{Z}}

\begin{document}
\title[Single radius spherical transform]{The support theorem for the single radius spherical mean transform}
\author{Mark Agranovsky}
\address{Mathematics Department\\ Bar Ilan University\\
Ramat Gan 52900
Israel}
\email{agranovs@macs.biu.ac.il}
\author{Peter Kuchment}
\address{Mathematics Department\\
Texas A\&M University\\
College Station, TX 77843-3368, USA}
\email{kuchment@math.tamu.edu}
\date{}
\subjclass[2000]{35L05, 92C55, 65R32, 44A12}
\keywords{Spherical mean, Radon transform, support}
\maketitle
\begin{abstract}
Let $f\in L^p(\R^n)$ and $R>0$. The transform is considered that integrates the function $f$ over (almost) all spheres of radius $R$ in $\R^n$. This operator is known to be non-injective (as one can see by taking Fourier transform). However, the counterexamples that can be easily constructed using Bessel functions of the 1st kind, only belong to $L^p$ if $p>2n/(n-1)$. It has been shown previously by S. Thangavelu that for $p$ not exceeding the critical number $2n/(n-1)$, the transform is indeed injective.

A support theorem that strengthens this injectivity result can be deduced from the results of \cite{Volch,Volch2}. Namely, if $K$ is a convex bounded domain in $\R^n$, the index $p$ is not above $2n/(n-1)$, and (almost) all the integrals  of $f$ over spheres of radius $R$ not intersecting $K$ are equal to zero, then $f$ is supported in the closure of the domain $K$.

In fact, convexity in this case is too strong a condition, and the result holds for any what we call $R$-convex domain.

We provide a simplified and self-contained proof of this statement.
\end{abstract}

\section{Introduction}

We consider the transform acting on functions defined on $\R^n$ by integrating them over all spheres of a fixed radius $R>0$. Let $f\in L^p(\R^n), 1\leq p\leq \infty$, then such integrals exist for almost every center. One can easily construct examples of non-injectivity of this transform at least for some values of $p$ (see the proof of Theorem \ref{T:main} below for details). However, such constructions, which use Bessel functions of the 1st kind, work only when $p>2n/(n-1)$. And indeed, it was shown by S. Thangavelu \cite{Thang} that for $p\leq 2n/(n-1)$, the transform is injective. In this text, we prove a stronger statement (comparable to S.~Helgason's ``hole'' support theorem \cite[Theorem 2.6 and Corollary 2.8]{Helg_Radon} for the Radon transform):

\begin{theorem}\label{T:main}
    Let $K$ be the closure of a bounded convex domain in $\RR^n$ with $n>1$, a function $f(x)$ belong to $L^p(\RR^n)$
    with $p\leq 2n/(n-1)$, and $R>0$. If the integrals of $f$ over almost all spheres of radius $R$ contained in $\RR^n\setminus K$ are equal to zero, then $f$ is compactly supported and its support is contained in $K$.

    This conclusion does not hold for $p> 2n/(n-1)$.
\end{theorem}

It is interesting to notice the appearance of the same critical power $2n/(n-1)$ in a similar situation, where however the set of spheres of integration is defined differently: one allows arbitrary radii of the spheres, but restricts the set of their centers to the points of a closed hypersurface $S\subset \R^n$ only. It is shown in \cite{ABK} that this transform is injective when $p\leq 2n/(n-1)$ and injectivity fails otherwise, for instance when $S$ itself is a sphere.

Convexity is too strong condition in this case. The statement holds for a larger class of domains that is natural for the problem under the consideration.

\begin{definition}\label{D:Rconvex}
Let $R$ be a positive number. A bounded closed domain $K\subset \R^n$ is said to be $R$-convex, if
\begin{enumerate}
  \item Its complement $CK:=\R^n\setminus K$ is the union of all closed balls $B\in CK$ of radius $R$.
  \item The set of centers of all such balls is connected.

\end{enumerate}
\end{definition}

\begin{theorem}\label{T:Rconvex}
    The statement of Theorem \ref{T:main} holds for $R$-convex bounded domains $K$.
\end{theorem}

Theorem \ref{T:main} is proven in the next section. In the following section, an auxiliary local result is established in Theorem \ref{T:local}. In the next section, Theorem \ref{T:Rconvex} is derived from Theorems \ref{T:main} and  \ref{T:local}. The paper ends with the remarks and acknowledgments sections.

A few month after the paper was posted on May 8th 2009, the book \cite{Volch2} appeared, which apparently contains results implying the theorems of this article. Moreover, it has been pointed out to us that these results can be derived from \cite[Chapter 3.3, Section 3.2, Corollary 3.3]{Volch}. It is, however, difficult to reconstruct the proof, which is distributed among various parts of the very technical books \cite{Volch,Volch2}. The authors thus think that a steamlined and self-contained proof would be useful to researchers in the area of integral geometry and harmonic analysis.

\section{Proof of Theorem \ref{T:main}}\label{S:proof}

We start with the hardest part of the proof, when $p\leq 2n/(n-1)$.

Let $f\in L^p(\RR^n),\quad p\in[1,2n/(n-1)]$ be such that
\begin{equation}\label{E:zeromeans}
   \int\limits_{|\omega|=1}f(y+R\omega)d\sigma(\omega)=0
\end{equation}
for almost all $y\in\RR^n$ such that $\mathrm{dist} (y,K) > R$, where $d\sigma(\omega)$ is the standard surface area measure on the unit sphere in $\R^n$. We need to show that then $f(x)=0$ for almost all $x\notin K$.

Since $K$, being a closed bounded convex domain, is the intersection of all balls it is contained within, it is sufficient to prove the statement when $K$ is a ball. Rescaling and shifting, we can assume without loss of generality that $K$ is the unit ball $B(0,1)$ centered at the origin.

Convolving with small support smooth radial functions, one reduces the problem to the case when $f$ is infinitely differentiable and, moreover, all its derivatives belong to the same space $L^p$ as $f$ itself.

Consider for each $m\in\ZZ^+$ an orthonormal basis  $Y^m_l, 1 \leq l \leq d(m)$ of the space of all spherical harmonics of degree $m$ in $\RR^n$ (the natural representation of the group $O(n)$ in this space is irreducible). Then function $f$ can be expanded into the Fourier series with respect to spherical harmonics as follows:
\begin{equation}\label{E:sph_series}
    f(x)=\sum\limits_{m,l}f_{m,l}(|x|)Y^m_l(\theta),
\end{equation}
where $\theta=\frac{x}{|x|}$ and
\begin{equation} \label{E:gml}
 f_{m,l}(|x|)=\int_{\theta \in S} f(|x|\theta) Y^m_l(\theta) d\theta.
\end{equation}

Due to the obvious rotational invariance of the problem, each term $f_{m,l}(|x|)Y^m_l(\theta)$ of the series also has the corresponding spherical integrals (\ref{E:zeromeans}) vanishing (see a more detailed consideration in Lemma \ref{L:harmonics} below). Since clearly $f_{m}$ belongs to the same $L^p$-space that $f$ does, it is sufficient to prove the statement of the theorem for the functions of the form
\begin{equation}\label{E:oneterm}
    f(|x|)Y^m_l(\theta)
\end{equation}
only, where, as before, $\theta=\frac{x}{|x|}$. Hence, we will assume from now on that $f$ is in the form (\ref{E:oneterm}).

Let $\delta_R (x)$ be the delta function supported on the sphere of radius $R$ centered at the origin. Then condition (\ref{E:zeromeans}) can be rewritten as follows:
\begin{equation}\label{E:convolve}
   h(x):=(f * \delta_R) (x)=0 \mbox{ for } |x|>R+1,
\end{equation}
where the star $*$ denotes the $n$-dimensional convolution. Considering $f(x)$ as a tempered distribution, one can pass to Fourier images in the left hand side of (\ref{E:convolve}) to get
\begin{equation}\label{E:convolveF}
   \widehat h(\xi)= \widehat{f}(\xi) \widehat{\delta_R} (\xi), \xi\in\RR^n_\xi.
\end{equation}
Notice that due to (\ref{E:convolve}), function $h:=f * \delta_R$ is compactly supported (with the support in the ball of radius $R+1$) and smooth, and thus the standard Paley-Wiener theorem applies \cite{Stein}. Therefore, the Fourier transform $\widehat{h}(\xi)$ of $h$ is an entire function satisfying for any $N>0$ the estimate:
\begin{equation}\label{E:PW}
|\widehat h(\xi)|\leq C_N (1+|\xi|)^{-N}e^{(R+1)|\Im \xi|}.
\end{equation}
We also recall that $\widehat{\delta_R} (\xi)$ coincides, up to a constant factor, with $j_{(n-2)/2} (R|\xi|)$, where
$j_p$  is the so called {\em normalized} or {\em spherical} Bessel function \cite{Magnus}:
\begin{equation}\label{E:modif_Bessel}
j_p (\lambda)=\frac{2^p\Gamma (p+1) J_p (\lambda)}{\lambda^p}.
\end{equation}
Here we use the standard notation $J_p(\lambda)$ for Bessel functions of the first kind.

Due to (\ref{E:convolveF}), we have
\begin{equation} \label{E:h(ksi)}
\widehat h(\xi)=\mathrm{const} j_{(n-2)/2}(R|\xi|)\widehat f(\xi).
\end{equation}

We can now explain the strategy of the proof. The claim we are proving is equivalent to $\widehat{f}(\xi)$ being an entire function of the following Paley-Wiener class:
\begin{equation}\label{E:PW2}
|\widehat f(\xi)|\leq C_N (1+|\xi|)^{-N}e^{R|\Im \xi|}
\end{equation}
(notice the exponent $R$ in (\ref{E:PW2}) instead of $R+1$ present in (\ref{E:PW})). Taking into account
(\ref{E:h(ksi)}), this task will be achieved, if we could show that:
\begin{enumerate}
\item The distribution $\widehat{f}$ does not have any delta-type terms supported at zeros of $j_{(n-2)/2}(R|\xi|)$, and thus  $\widehat{f}$ can be obtained by dividing $\widehat{h}$ by $j_{(n-2)/2}(R|\xi|)$.

\item This ratio is entire, i.e. $\widehat{h}$ in fact vanishes at zeros of $j_{(n-2)/2}(R|\xi|)$.

\item The estimate (\ref{E:PW2}) holds, which due to (\ref{E:PW}) requires one to get an estimate from below for $j_{(n-2)/2}(R|\xi|)$ that would eliminate the unnecessary $+1$ in $R+1$ in (\ref{E:PW}).
\end{enumerate}

We will deal with these steps in the reverse order. The last one is achieved by the
following simple statement:
\begin{lemma}\label{L:bessel}(e.g., \cite[Lemma 6]{AmbKuc_range} or \cite[Lemma 4]{AFK})
On the entire complex plane, except for a disk $S_0$ centered at
the origin and a countable number of disks $S_k$ of radii $\pi/6$
centered at points $\pi(k+\frac{2\nu+3}{4})$, one has
\begin{equation}
\label{E:estimate} |J_\nu (z)|\ge\frac{C
e^{|Im\,z|}}{\sqrt{|z|}},\quad C>0.
\end{equation}
\end {lemma}

In order to handle the other two issues, we need to do some preparations.

The following lemma allows one to represent spherical means as volume integrals.

\begin{lemma}\label{L:represent}
Let $\lambda_0>0$ satisfy
$$ j_{(n-2)/2}(R\lambda_0)=0.$$
Then the spherical mean  $h=\delta_R * f$ can be represented as
\begin{equation}\label{E:representation}
 h={\mathrm const} (\Delta +\lambda_0^2)(f * \Psi_R)
\end{equation}
where
$$\Psi(x)= j_{(n-2)/2}(\lambda_0|x|)\chi_R(x)$$
and $\chi_R$ is the characteristic function of the ball of radius $R$ centered at the origin.
\end{lemma}
\begin{proof}
Indeed, this follows easily from Stokes formula. Denoting by $B(x,R)$ and $S(x,R)$ the ball and sphere centered at $x$ and of radius $R$, one gets
\begin{equation}\label{E:stokes}
\begin{array}{c}
\int\limits_{B(x,R)}\{[(\Delta+\lambda_0^2)f](v)j_{(n-2)/2}(\lambda_0|x-v|)\\
-f(v)
[(\Delta+\lambda_0^2)j_{(n-2)/2}](\lambda_0|x-v|)\}dv \\
= \int\limits_{S(x,R)} \{f(v)\frac{dj_{(n-2)/2}}{dr}(R\lambda_0)-
\frac{df}{dr}(v)j_{(n-2)/2}(R \lambda_0)\}dA(v).
\end{array}
\end{equation}
Here $r=|x|$ and $\partial/\partial r$ is the external normal derivative on the sphere $|x|=t$.

We now take into account that, according to our choice of $\lambda_0$, the Bessel function $j_{(n-2)/2}(\lambda_0 u)$ satisfies the following two equalities:
$$
j_{(n-2)/2}(R\lambda_0)=0
$$
and
$$
(\Delta+\lambda_0^2)j_{(n-2)/2}(\lambda_0 |y|)=0.
$$
Also, due to the simplicity of zeros of $j_{(n-2)/2}$,
$$
j^\prime_{(n-2)/2}(R\lambda_0) \neq 0.
$$
These features, combined with (\ref{E:stokes}), prove the statement of the lemma.
\end{proof}

\begin{lemma}\label{L:harmonics}
Let $f(x)=\sum\limits_{l=1}^{d(m)}f_l(r)Y^m_l(\theta), \ x=r\theta, |\theta|=1.$ Then for any radial compactly supported continuous
function $\psi$ the convolution $F=\psi * f$ has the similar representation $F(x)=\sum\limits_{l=1}^{d(m)}F_l(r)Y^m_l(\theta).$
\end{lemma}
\begin{proof} The convolution operator $f\to \psi*f$ is rotationally invariant. Indeed:
$$
(\psi*f)(x)=\int f(y)\psi(|x-y|))dy,
$$
and thus for any rotation $T$ and the rotated function $f_T(x)=f(Tx)$ one has:
$$
\begin{array}{c}
(\psi*f_T)(x)=\int f(Ty)\psi(|x-y|))dy=\int f(Ty)\psi(|Tx-Ty|))dy\\
=\int f(y)\psi(|Tx-y|))dy=(\psi*f)_T(x).
\end{array}
$$
This implies that the convolution preserves the subspaces of harmonics of a fixed degree, which proves the lemma.
\end{proof}

{\bf End of the proof of Theorem 1}

According to our strategy, the next step is to prove that $\widehat f$ is an entire function.

Due to (\ref{E:h(ksi)}), outside of zeros of $j_{(n-2)/2}(R|\xi|)$, one has
\begin{equation}\label{E:ratio}
\widehat f(\xi)=const \frac{\widehat h(\xi)}{j_{(n-2)/2}(R|\xi|)}.
\end{equation}
Notice that the denominator is an entire function of the variable $\xi \in \mathbb C^n$, since
$j_\nu(u)$ is an even entire function of the real argument $u$ and hence is an entire function of  $u^2$
The next lemma shows that the numerator in (\ref{E:ratio}) vanishes at the (simple) zeros of the denominator.
Therefore, the zeros cancel, and the ratio in the right hand side of (\ref{E:ratio}) is an entire function, as needed.

\begin{lemma}\label{L:zeros}
 For any $\lambda_0$ such that $j_{(n-2)/2}(R\lambda_0)=0$, function $\widehat h(\xi)$ vanishes
on the complex quadric
$$
Q=\{\xi \in \mathbb C^n \ | \ \xi_1^2+...\xi_n^2=\lambda_0^2 \}.
$$
\end{lemma}
\begin{proof}
Since $\lambda_0\neq 0$, the quadric $Q$ is irreducible and has a maximal dimension intersection with the real subspace. Thus, due to analytic continuation, it suffices to check vanishing of the entire function $\widehat h(\xi)$ on the
intersection $Q\bigcap \R^n$, i.e. on the sphere $|\xi|=\lambda_0$ in $\mathbb R^n$.

Since, by assumption, $h$ vanishes outside of the unit ball, we can write
$$
\widehat h(\xi)=\int\limits_{|x| \leq t }h(x) e^{-i \xi \cdot x}dx,
$$
for arbitrary $t>1$.

Let us substitute for $h$ the representation (\ref{E:representation}). Then, by Stokes' formula,
\begin{equation}
\begin{array}{c}
\widehat h(\xi)={\mathrm const} \int\limits_{|x| \leq t}(\Delta +\lambda_0^2)(f * \Psi_R) e^{-i \xi \cdot x}dx\\=
{\mathrm const} \int\limits_{|x| \leq t} (f *\Psi_R)(\Delta +\lambda_0^2)e^{-i \xi \cdot x} dx\\
+{\mathrm const}\int\limits_{|x|=t}(\frac{\partial}{\partial r}(f *\Psi_R)e^{-i \xi \cdot x}-
(f *\Psi_R)\frac{\partial}{\partial r} e^{-i \xi \cdot x}) dA(x).
\end{array}
\end{equation}
Since $|\xi|=\lambda_0$, the exponential function $e^{-i \xi \cdot x}$ is annihilated by the operator $\Delta +\lambda_0^2$.
Therefore, $\widehat h(\xi)$ is expressed by the surface term alone:
$$
\widehat h(\xi)=const \int\limits_{|x|=t}(\frac{\partial}{\partial r}(f * \Psi_R)e^{-i \xi \cdot x}-
(f *\Psi_R) \frac{\partial}{\partial r} e^{-i \xi \cdot x}) dA(x).
$$
Here, as before, $r=|x|$ and $\frac{\partial}{\partial r}$ is the external normal derivative on the sphere $|x|=t$.

The function $\Psi_R$ is radial and thus, due to Lemma \ref{L:harmonics}, the convolution
$F:=f * \Psi$ has the form $F(x)=\sum_{l=1}^{d(m)}F_l(r)Y_l(\theta)$.
Projection of the exponential function $e^{-i \xi \cdot x}$ on the space of spherical harmonics of degree $m$ can be given in terms of Bessel functions (see \cite[Theorem 3.10]{Stein}), which leads to the following formula:
\begin{equation}\label{E:h_at_M}
\widehat h(\xi)=c_m\lambda_0^m t^{n+m-1}\sum\limits_{l=1}^{d(m)} \left(F_l^{\prime}(t) j_{n/2+m-1}(\lambda_0 t)-
F_l(t) j^{\prime}_{n/2+m-1}(\lambda_0 t)\right).
\end{equation}
In what follows, the estimate is done the same way for any $l$ between $1$ and $d(m)$, so we will drop the sum over $l$ and work with a single term.

In order to prove that $\widehat{h}(\xi)=0$, it suffices to check that the expression in the right hand side tends to 0 as $t \to \infty$. This can now be easily shown using the $L^p$ condition on $F$ and the known estimate for Bessel functions:
\begin{equation}\label{E:besselestimate}
j_{n/2+m-1}(t), j^{\prime}_{n/2+m-1}(t)=O(t^{-\frac{n+2m-1}{2}}), \ t \to \infty.
\end{equation}
 Indeed, let us pick $t_0>t$ and average both sides of (\ref{E:h_at_M})
for $t$ from $t_0$ to $2t_0$:
\begin{equation}\label{E:estimate}
\begin{array}{c}
\widehat h(\xi)=c_m \frac{1}{t_0}\int\limits_{t_0}^{2t_0} [F_l^{\prime}(t) j_{n/2+l-1}(\lambda_0 t)\\
- F_l(t) j^{\prime}_{n/2+l-1}(\lambda_0 t)]t^{n+m-1}dt.
\end{array}
\end{equation}
Let $A(t):=|F_l^{\prime}(t)|+|F_l(t)|$. From (\ref{E:besselestimate}) and (\ref{E:estimate}) one obtains:
\begin{equation}\label{E:M}
|\widehat h(\xi)| \leq \frac{c_m}{t_0}\int\limits_{t_0}^{2t_0}A(t)t^{\frac{n-1}{2}}dt
=\frac{c_m}{t_0}\int\limits_{t_0}^{2t_0}A(t)t^{\frac{n-1}{p}}t^{(n-1)\frac{p-2}{2p}}dt.
\end{equation}

Functions $F_l(r)$ and $F_l^{\prime}(r)$ are the radial parts of functions in $L^p(\R^n)$ and therefore
belong to $L^p((0,\infty),r^{n-1}dr)$. So is the function $A(r)$.
We now apply H\"{o}lder inequality to (\ref{E:M}) to get
\begin{equation}\label{E:Holder}
|\widehat h(\xi)| \leq \frac{c_m}{t_0}\left(\int\limits_{t_0}^{2t_0}A^p(t)t^{n-1}dt\right)^{\frac{1}{p}}
\left(\int\limits_{t_0}^{2t_0}t^{(\frac{n-1}{2}-\frac{n-1}{p})q}dt\right)^{\frac{1}{q}},
\end{equation}
where the index $q$ dual to $p$ is introduced in the standard manner: $p^{-1}+q^{-1}=1$, or $q=p/(p-1)$.
The second factor in (\ref{E:Holder}) can be easily computed:
$$
\left(\int_{t_0}^{2t_0}t^{(\frac{n-1}{2}-\frac{n-1}{p})q}dt\right)^{\frac{1}{q}}
=Ct_0^{\frac{n-1}{2}-\frac{n}{p}+1},
$$
and hence (\ref{E:Holder}) leads to the estimate:
\begin{equation}\label{E:estimate2}
|\widehat h(\xi)| \leq c_m\|A\|_{L^p((t_0,2t_0), t^{n-1}dt)} t_0^{\frac{n-1}{2}-\frac{n}{p}}.
\end{equation}
The condition $p\leq 2n/(n-1)$ shows that $(n-1)/2-n/p \leq 0$, and hence the last factor in (\ref{E:estimate2}) is bounded. Since the condition that $F\in L^p(\R^n)$ implies
$$
\|A\|_{L^p((t_0,2t_0), t^{n-1}dt)} \to 0 \mbox{ when } t_0 \to \infty,
$$
this shows the required equality $\widehat h(\xi)=0$.
\end{proof}

\begin{corollary}\label{C:PW}
The function
$$
\Phi(\xi):=\frac{\widehat{h}(\xi)}{j_{(n-2)/2}(R|\xi|)}
$$
is entire of the Paley-Wiener class (\ref{E:PW2}).
\end{corollary}

The only remaining step is to show that the same statement as in Corollary \ref{C:PW} applies to the function $\widehat f(\xi)$:
\begin{lemma}\label{E:entire}
The Fourier transform $\widehat f (\xi)$ is an entire function of the Paley-Wiener class (\ref{E:PW2}).
\end{lemma}
\begin{proof}
Corollary \ref{C:PW} says that the right hand side in (\ref{E:ratio})
is an entire function  of the Paley-Wiener class (\ref{E:PW2}). The Lemma (and thus the Theorem \ref{T:main}) will be proven if we show that in fact $\widehat f=\Phi$.

The tempered distribution $\widehat f (\xi), \xi \in \mathbb R^n$
coincides with $\Phi(\xi)$ outside of the union of the discrete set of spheres $S_k$ defined by
(simple) zeros of Bessel function:
$$
S_k=\{\xi \in \mathbb R^n: \xi_1^2+...+\xi_n^2=\lambda_k^2\},
$$
where
$$
j_{(n-2)/2}(\lambda_k R)=0.
$$
This means that $\widehat f$ can differ from $\Phi$ only by terms supported on these spheres:
$$
\widehat f (\xi)=\Phi(\xi) + \sum_k c_k(\xi) \delta(|\xi|- |\lambda_k|).
$$
Although in principle higher order distibituions concentrated on the spheres could have arised, the equality (\ref{E:h(ksi)}), together with the simplicity of zeros of the Bessel function involved, shows that these higher order terms are not present.
 
We now observe that since $f(x)=\sum_{l=1}^{d(m)}f_l(r)Y_l^m(\theta)$, the coefficients $c_k(\xi)$ must have the similar form
$$
c_k(\xi)=\sum_{l=1}^{d(m)}a_{k,l} Y_l^m(\eta),  a_k=const, \xi=|\xi|\eta, |\eta|=1.
$$
Our aim is to show that there are no such distributional terms in $\widehat f$, i.e. all coefficients $a_{k,l}$ must vanish.

Fix $k$ and choose a positive number $\varepsilon$
so small that the spherical layer
$$
L:=\{\lambda_k-\varepsilon \leq |\xi| \leq \lambda_k +\varepsilon\}
$$
containing $S_k$, does not contain other spheres $S_m$ with $m \neq k$.

Let now $\psi$ be a radial function from the Schwartz class, whose Fourier transform vanishes outside the
spherical layer $L$ and such that $\widehat\psi (\xi)=1, \ \xi \in S_k$. We can now localize the sphere $S_k$ in the spectrum of $f$ by considering the convolution $g=\psi * f$.
By construction,
$$
\widehat g(\xi)=\Phi(\xi) \widehat \varphi (\xi) + c_kY_l(\eta) (\delta(|\xi|-\lambda_k).
$$
The first term is in the Schwartz class, while the second one is, up to a constant factor, Fourier transform of Bessel function $j_{n/2+l-1}(|x|)$
and therefore after convolving with $\psi$ we have
$$g(x)= \psi * \varphi + const \ a_k j_{n/2+l-1}(|x|),$$
where $\psi$ is inverse Fourier transform of $\Psi$ and hence is also a Schwartz function.
By the condition for $f$ and by the construction, the functions $g$ and $\psi *\varphi $ belong to $L^p(\mathbb R^n)$ with
$p<2n/n-1$, while Bessel function $j_{n/2+l-1}$ is not in this class. Therefore the coefficient $a_k$ must be equal to zero.

Thus, there is no $\delta$-function terms in $\widehat f$ and
$\widehat f=\Psi$ is an entire function in $\mathbb C^n$ satisfying, as it was explained above,
the Paley-Wiener estimate that implies that $supp f \subset \overline B(0,R)$.

Let now $p> 2n/(n-1)$. Then one can find a counterexample, where even compactness of support of $f$ cannot be guaranteed, using Bessel functions. The function
\begin{equation}\label{E:f}
    f(x)=|x|^{1-n/2}J_{n/2-1}(\lambda|x|)
\end{equation}
provides such a counterexample (due to L.~Zalcman). Indeed, consider the following spherical mean mapping $M$:
$$
Mg(x,t)=\frac{1}{\omega_n}\int\limits_{S(0,1)} g(x+t\theta)d\theta,
$$
which averages any continuous function $g$ over spheres. It is well known that $f$ defined in \ref{E:f} satisfies the following functional identity:
\begin{equation}\label{E:relat}
Mf(x,t)=\mbox{const } f(x)f(t).
\end{equation}
Thus, if $\lambda$ is chosen as a zero of $J_{n/2-1}$, the relation (\ref{E:relat}) implies that the spherical means of $f(x)$ over all spheres of radius $1$ are equal to zero. Also, the known asymptotic behavior of Bessel functions shows that $f\in L^q(\R^n)$ for any $q>2n/(n-1)$.
This completes the proof of Theorem \ref{T:main}.
\end{proof}

\section{A local result}\label{S:local}

In order to extend the statement of Theorem \ref{T:main} to all $R$-convex domains, we need to establish first the following local theorem, which in some particular cases as well as in different related versions has been established previously \cite{John,Volch}.

\begin{theorem}\label{T:local}
   Let $f(x)$ be an infinitely differentiable function in the ball $B(0,R+\varepsilon)\subset \R^n$ and its spherical averages over all spheres of radius $R$ contained in this ball are equal to zero. If $f$ vanishes in the ball $B(0,R)$, then it vanishes in the whole ball $B(0,R+\varepsilon)$.
\end{theorem}
\begin{proof}
Without loss of generality, we can assume that $R=1$.
As in \cite{John,Volch}, we will exploit relations between spherical and plane waves \cite[Ch. 1 and 4]{John}.

For a function $u(x)$ on $\R^n$ we will denote by $u^\#(x)$ its {\em radialization}
$$
u^\#(x):=\int\limits_{k\in O(n)}u(kx)dk,
$$
where $dk$ is the normalized Haar measure on $O(n)$.
Function $u^\#(x)$ is clearly radial and thus is a function of a single variable $|x|$. Abusing notations, we will write $u^\#(x)=u^\#(|x|)$.

The following simple statement (which we will prove for completeness) will be useful:
\begin{lemma}\label{L:radialization}
 Let $u(x), v(x)$ be continuous functions on $\R^n$ and $v(x)$ be radial and compactly supported. Then
\begin{equation*}
(u\ast v)^\#=u^\# \ast v.
\end{equation*}
\end{lemma}
\begin{proof} Indeed,
\begin{equation*}
(u\ast v)^\#(x)=\int\limits_{O(n)}\int\limits_{\R^n}u(kx-y)v(y)dydk.
\end{equation*}
Changing the variables in the $y$-integral from $y$ to $ky$, using the rotational invariance of $v$, and changing the order of integration, one gets
\begin{equation*}
 (u\ast v)^\#(x)=\int\limits_{\R^n}
\left(\int\limits_{O(n)}u(kx-ky)dk\right)v(y)dy=(u^\# \ast v)(x).
\end{equation*}
This proves the lemma.
\end{proof}
In particular, the convolution of two radial functions is radial.

The relation between plane waves and radial functions that we need is contained in the following result of \cite[Ch.4, formulas (4.13) and (4.16)]{John}:

\begin{lemma}\cite{John}\label{L:John} Let $e\in\R^n$ and $g(p)$ be a function of a scalar variable $p\in\R$. We consider the ridge function $g(\langle x, e \rangle)$ and its radialization
$g(\langle \cdot, e \rangle)^\#$, which we will identify with a function $f(r)$ of scalar variable $r$.
Then the relations between the functions $f(r)$ and $g(p)$ are provided by the following Abel type transforms:

\begin{equation}\label{E:Abel}
f(r)=(\mathcal Ag)(r):=2\frac{\omega_{n-1}}{\omega_n} r^{2-n} \int\limits_0^r(r^2-s^2)^{\frac{n-3}{2}}g(p)dp
\end{equation}
and
\begin{equation}\label{E:Abel-inv}
g(p)=(\mathcal A^{-1}f)(p):=
\frac{2^{n-1}p}{(n-2)!}\left(\frac{d}{dp^2}\right)^{n-1}\int_0^p r^{n-1}(p^2-r^2)^{\frac{n-3}{2}}f(r)dr.
\end{equation}
\end{lemma}

We can now derive the following useful relation:

\begin{lemma}\label{L:Abel-conv}
Let $\delta_S$ denote the normalized measure supported by the unit sphere. Let also $g(p)$ be a continuous function on $\R$. 
Then
\begin{equation}\label{E:Abel-con}
 (\mathcal A g *\delta_S)(p)=\mathrm{const} \mathcal A( g *_{1} (1-|p|^2)_+^{\frac{n-3}{2}}),
\end{equation}
where $*_{1}$ denotes one-dimensional convolution and $\mathcal A g$ in the left hand side is considered as a radial function on $\R^n$, i.e. $\mathcal A g (|x|)$ for $x\in\R^n$.
\end{lemma}
\begin{proof}
Since $\mathcal A g=g(\langle \cdot, e\rangle)^\#$, the left hand side, according to Lemma \ref{L:radialization} can be rewritten as
$$
\left(g(\langle \cdot, e\rangle) * \delta_S\right)^\#.
$$
It is straightforward to check that
$$
\left(g(\langle \cdot, e\rangle) * \delta_S\right)(x)
$$
is equal to the ridge function
$$
( g *_{1} (1-|p|^2)_+^{\frac{n-3}{2}})|_{p=\langle x,e\rangle}.
$$
Now radialization of this ridge function gives the right hand side expression in (\ref{E:Abel-con}).
\end{proof}

We can complete now the proof of our theorem.
We start with the case of a radial function, which we write as $f(|x|)$ for some function $f(r)$ of a single variable.
By the assumption, $(f *\delta_S)(x)=0$ for $|x|<\varepsilon$. Then (\ref{E:Abel-con}) implies
that
\begin{equation}\label{E:epsilon}
(g *_{\mathbb R^1} (1-|p|^2)^{\frac{n-3}{2}})(s)=0
\end{equation}
for $ s \le \varepsilon,$ where $g(p): =(\mathcal A^{-1}f)(p).$
It follows from (\ref{E:Abel-inv}) that the condition
$f(x)=0$ for $|x| \leq 1$ implies  $g(p)=0$ for $|p| \leq 1$,
and therefore ( \ref{E:epsilon}) can be rewritten as
$$
\int\limits_1^{1+\varepsilon}
g(p)(1-|p-s|^2)_+^
{\frac{n-3}{2}}dp=0, s \leq \varepsilon.
$$
Thus the Titchmarsh theorem \cite{Titch} (see also \cite[Theorem 4.3.3]{Horm}, \cite[Lecture 16]{Levin}, or \cite[Ch. VI]{Yosida}) implies that $g(p)=0$ for $1\leq p \leq 1+\varepsilon$. Since $f=\mathcal A g$, the relation (\ref{E:Abel}) leads to the conclusion that $f(x)=0$ for $|x| \leq 1+\varepsilon$. This proves the statement of the theorem in the radial case.

It remains now to pass from radial to non-radial functions. To this end, we observe that the $C^{\infty}$  function $f$ has zero integrals over all spheres of radius $1$ centered in the open ball $B(0,\varepsilon)$. Thus, all its partial derivatives $D^{\alpha}f$ have the same property.
Since this vanishing condition is invariant under rotations, according to Lemma \ref{L:radialization}, it also holds for radializations $(D^{\alpha}f)^{\#}$. Since the theorem is already proven for radial functions,
all these radializations vanish, i.e.
\begin{equation}\label{E:vanish}
\int\limits_{|x|=r}D^{\alpha}f(x)dA(x)=0
\end{equation}
for all $0 < r < 1+\varepsilon$.

Let us prove now that on each sphere $|x|=t$ for $t\in[0,1+\varepsilon)$ the function $f$, along with all its derivatives, is orthogonal to all monomials. This, due to the Weierstrass Theorem will imply the needed property that $f=0$ in $B(0,1+\varepsilon)$.

We prove this claim by induction with respect to the degree of the monomial. For a zero degree monomial, the claim is true, due to (\ref{E:vanish}). Suppose that
\begin{equation}\label{E:induction}
\int_{|x|=t}p(x)D^{\alpha}f(x)dA(x)=0,
\end{equation}
 for all  monomials $p(x)$ of degree not exceeding $N$ and all multiindices $\alpha$.
Integrating both sides of this identity with respect to $t$ from $0$ to any $r <1+\varepsilon$ yields
$$
\int\limits_{|x|\leq r} p(x)D^{\alpha}f(x)dx=0.
$$
We now replace the multiindex $\alpha$ with $\beta=\alpha +\delta_j$, where the multiindex $\delta_j$ has 1 in $j$th place and $0$s otherwise and write
\begin{equation} \label{E:p(x)}
p(x)D^{\beta}f(x)=\frac{\partial}{\partial x_j} \left(p(x)D^{\alpha}f(x)\right)-
\frac{\partial p}{\partial x_j} D^{\alpha}f(x).
\end{equation}
The second term on the right does not contribute to the integral over the ball $|x| \leq r$, due to the induction assumption, and thus identity (\ref{E:induction}), where $\alpha$ is replaced by $\beta$, reduces to
$$
\int\limits_{|x| \leq r}\frac{\partial}{\partial x_j}(p(x)D^{\alpha}f(x)) dx=0.
$$
Using Stokes' formula, we obtain
$$
\int\limits_{|x|=r} x_j p(x) D^{\alpha}f(x)dA(x)=0.
$$
Since $j=1,\dots,n$ is arbitrary, we conclude that  identity (\ref{E:induction}) holds for all monomials of degree $N+1$. This completes the proof of theorem.
\end{proof}

\section{Proof of Theorem \ref{T:Rconvex}}\label{S:Rconvex}

We can now prove Theorem \ref{T:Rconvex} that extends Theorem \ref{T:main} to the case of $R$-convex domains. So, we assume that $K\subset\R^n$ is a closed $R$-convex domain and a function $f\in L^p(\R^n)$ with $p\leq 2n/(n-1)$ is such that its spherical means over almost every sphere of radius $R$ not intersecting $K$ is zero. As it has been shown before, one can assume, without restriction of generality, that the function is smooth. Consider the set $C$ of centers of all balls of radius $K$ not intersecting $K$. Due to $R$-convexity of $K$, this set is connected, and the union of the corresponding balls covers the whole complement of $K$. Consider also the subset $C_f\subset C$ of such centers $x$ that $f$ vanishes in the ball $B(x,R)$. If we establish that in fact $C_f=C$, this will prove the theorem.

Theorem \ref{T:main} implies that $f=0$ outside the convex hull of $K$. Thus, in particular, the set $C_f$ is non-empty, since it contains all points $x$ with a sufficiently large norm. It is also obvious that, due to continuity of $f$, the set $C_f$ is relatively closed in $C$. Let us now prove that it is also relatively open. Due to connectedness of $C$, this will imply that $C_f=C$ and thus $f=0$ in the whole complement of $K$, which is the statement of the theorem.

Indeed, let $x\in C_f$. This means that $f=0$ in $B(x,R)$. There exists a positive $\varepsilon$ such that the ball $B(x,R+\varepsilon)$ is inside the complement of $K$. Then the function $f$ satisfies the conditions of Theorem \ref{T:local} in $B(x,R+\varepsilon)$, and thus $f=0$ in $B(x,R+\varepsilon)$. In particular, $f$ vanishes in the ball $B(y,R)$ for any $y\in\R^n$ such that $|y-x|<\varepsilon$. This means that all such points $y$ belong to $C_f$, and hence $C_f$ is open. This finishes the proof of the theorem.
\qed
\section{Remarks}\label{S:remarks}

\begin{enumerate}
\item As it was mentioned in the introduction, the results contained in V.~Volchkov's books \cite{Volch,Volch2} imply Theorem \ref{T:main}. It is, however, not that easy to reconstruct its proof, which is distributed among various parts of \cite{Volch,Volch2}. The authors thus think that a steamlined proof and self-contained would be useful to researchers in the area of integral geometry and harmonic analysis.

\item The statement of Theorem \ref{T:local} holds also for functions of finite smoothness, if one knows that spherical averages of $f$ vanish for all spheres of radius $r<R$ (rather than $r=R$ as in Theorem \ref{T:local}).

\item The local Theorem \ref{T:local},  has been established previously in some particular cases, as well as in different related versions in \cite{John,Volch}. For instance, one can check that the consideration in the second section of \cite[Ch. VI]{John} provides such a result in $3D$, although the local formulation is not stated there. In \cite{Volch}, a theorem similar to Theorem \ref{T:local} is proven for the case of integrals over balls.
\end{enumerate}
\section*{Acknowledgments}
The work of the first author was performed when he was visiting Texas A\&M University and was partially supported by the ISF (Israel Science Foundation) grant 688/08. The second author was partially supported by the NSF grants DMS 0604778 and 0908208 and by the IAMCS. The authors express their gratitude to ISF, NSF, Texas A\&M University, and IAMCS for the support. We are also grateful to the reviewer of the first version of the text for pointing to us the reference to \cite{Volch2}.

\end{document}